\documentclass{elsart}

\usepackage{epsfig}

\usepackage{amssymb}

\begin{document}

\begin{frontmatter}



\title{Electrodynamical properties of a "grid" volume resonator for travelling wave tube and backward wave oscillator}


\author{V. Baryshevsky},
\author{A. Gurinovich}

\address{Research Institute of Nuclear Problem,
Belarussian State University,   \\   11 Bobruyskaya Str. , Minsk 220050, Belarus}

\begin{abstract}
The electrodynamical properties of a volume resonator formed by a
perodic structure built from the metallic threads inside a
rectangular waveguide ("grid" volume resonator) is considered for
travelling wave tube and backward wave oscillator operation.
Peculiarities of passing of electromagnetic waves with different
polarizations through such volume resonator are discussed.
\end{abstract}


\end{frontmatter}

\qquad
\section{Introduction}
Generation of radiation in millimeter and far-infrared
range with nonrelativistic and low-relativistic electron beams
gives rise difficulties. Gyrotrons and cyclotron resonance
facilities are used as sources in millimeter and sub-millimeter
range, but for their operation magnetic field about several tens
of kiloGauss ($\omega \sim \frac{eH}{mc}\gamma $) is necessary.
Slow-wave devices (TWT, BWT, orotrons)in this range require
application of dense and thin ($<0.1$ mm) electron beams, because
only electrons passing near the slowing structure at the distance
$\leq \lambda \beta \gamma /(4\pi )$ can interact with
electromagnetic wave effectively.
It is difficult to guide thin beams near slowing structure with
desired accuracy. And electrical endurance of resonator limits
radiation power and density of acceptable electron beam.
Conventional waveguide systems are essentially restricted by the
requirement for transverse dimensions of resonator, which should
not significantly exceed radiation wavelength. Otherwise,
generation efficiency decreases abruptly due to excitation of
plenty of modes. The most of the above problems can be overpassed
in VFEL
\cite{PhysLett,VFELreview,FirstLasing,FEL2002,patent}.
In VFEL the greater part of electron beam interacts with
the electromagnetic wave due to volume distributed interaction.
Transverse dimensions of VFEL resonator could significantly exceed
radiation wavelength $D \gg \lambda $. In addition, electron beam
and radiation power are distributed over the whole volume that is
beneficial for electrical endurance of the system. Multi-wave
Bragg
dynamical diffraction provides mode discrimination in VFEL.

The electrodynamical properties of volume
diffraction structures composed from strained dielectric threads was experimentally
studied in
\cite{VolumeGrating}. In \cite{THz} it was shown that nonrelativistic and
low-relativistic electron beams passing through such structures
can generate in wide frequency range up to terahertz.

\begin{figure}[h]
\epsfxsize = 8 cm
\centerline{\epsfbox{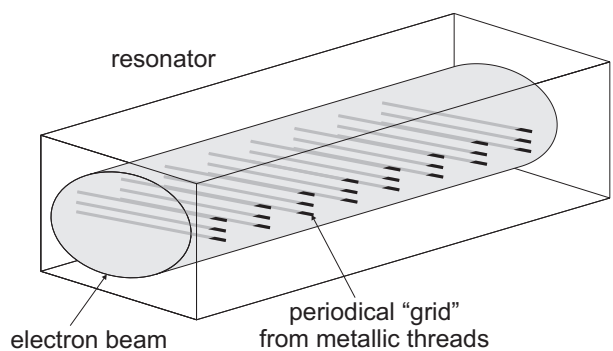}}
\caption{"Grid" volume resonator}
\label{resonator}
\end{figure}

\qquad
In the present paper the electrodynamical properties of a "grid" volume resonator
that is formed by a perodic structure built
from the metallic threads inside a rectangular waveguide (see Fig.\ref{resonator})
is considered.

\section{Scattering by a set of metallic threads}

\begin{figure}[h]
\epsfxsize = 8 cm
\centerline{\epsfbox{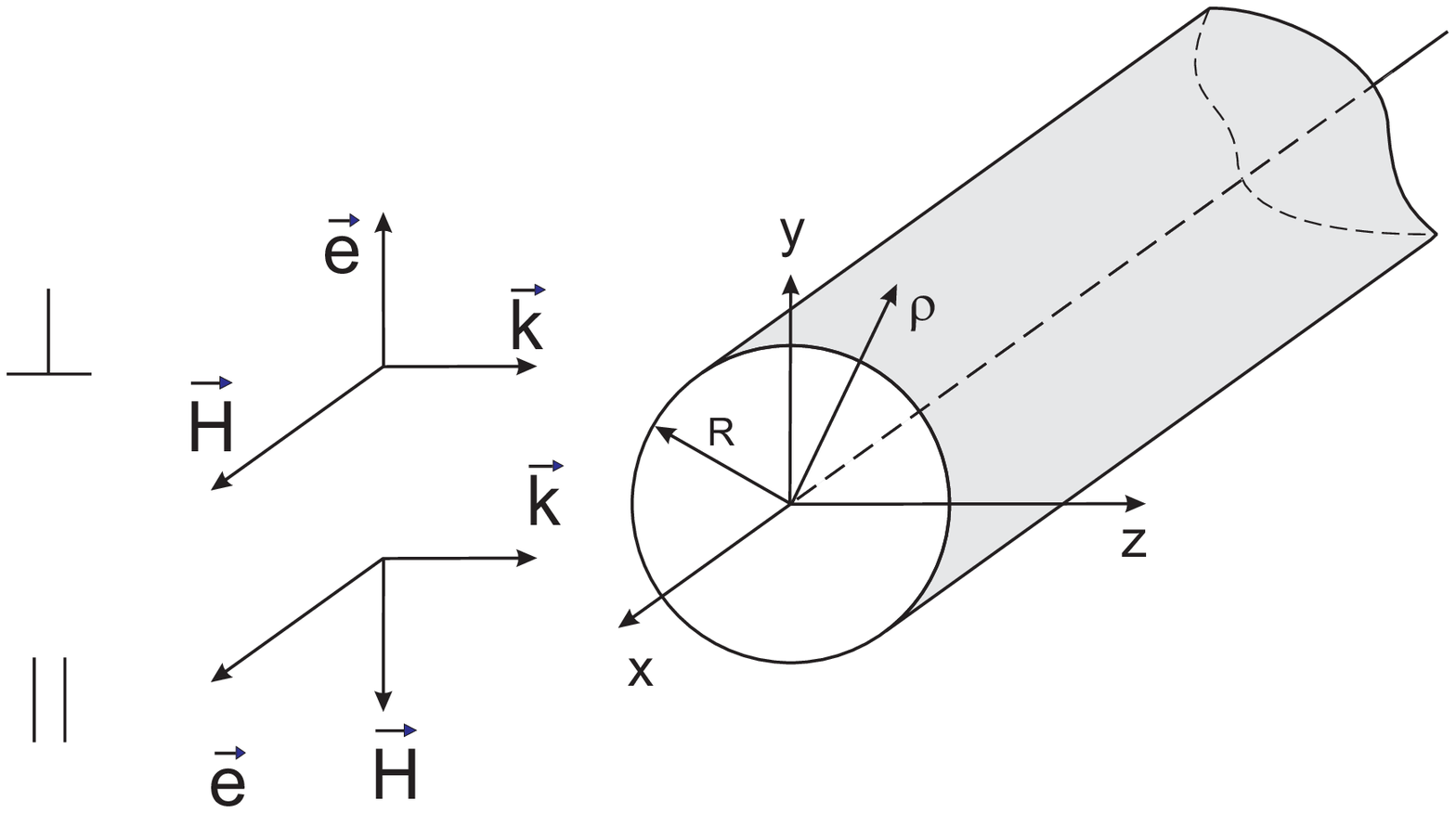}}
\caption{.}
\label{cylinder}
\end{figure}

Let us consider a plane electromagnetic wave $\overrightarrow{E}=\Psi \overrightarrow{e}$,
where $\overrightarrow{e}$ is the polarisation vector.
Suppose this wave
falls onto
the cylinder placed into the origin of coordinates and the cylinder axis
concides with the axis $x$ (Fig.\ref{cylinder}) (in further consideration
$\overrightarrow{e}$ will be omitted). Two orientations of $\overrightarrow{e}$
should be considered: $\overrightarrow{e}$ is parallel to the cylinder axis $x$
and $\overrightarrow{e}$ is perpendicular to the cylinder axis $x$.
For clarity suppose that $\overrightarrow{e} \parallel 0x$.

The scattered wave can be written as \cite{Nikolsky}
\begin{equation}
\Psi=e^{ikz}+a_0 H_0^{(1)}(k \rho)
\label{1cylinder}
\end{equation}
here $\rho$ is the transverse coordinate $\rho=(y,z)$, $H_0^{(1)}$ is the Hankel function

Thus, considering a set of cylinders with $\rho_n=(y_n,z_n)$ one can express the scattered
wave as
\begin{equation}
\Psi=e^{ikz}+a_0 \Sigma_{n} H_0^{(1)}
(k \left\vert \overrightarrow{\rho }-\overrightarrow{\rho }_{n}\right\vert)e^{ikz_n}
\label{sum_cylinder}
\end{equation}
or using the integral representation for Hankel functions
\begin{equation}
\Psi=e^{ikz}+A_0 \Sigma_{n}
\int\limits_{-\infty }^{\infty }\frac{e^{i k\sqrt{\left\vert \overrightarrow{%
\rho }-\overrightarrow{\rho }_{n}\right\vert ^{2}-x^{2}}}}{\sqrt{\left\vert
\overrightarrow{\rho }-\overrightarrow{\rho }_{n}\right\vert ^{2}-x^{2}}}dx
e^{ikz_n},
\label{ncylinder}
\end{equation}
where $A_0=-\frac{i a_0}{\pi}$, $\left\vert
\overrightarrow{\rho }-\overrightarrow{\rho }_{n}\right\vert ^{2}=(y-y_n)^2+(z-z_n)^2$.

Let us consider the wave passing through the layer of cylinders,
which axes are distributed in the plane $x0y$
on the distance $d_y$. Summation over the coordinates $y_n$ provides the
following expression for $\Psi$:
\begin{equation}
\Psi=e^{ikz}+\frac{2 \pi i A_0}{k d_y}e^{ikz},
\label{average-n-cylinder}
\end{equation}
Thus, after passing m planes (standing out of each other in the distance $d_z$)
the scattered wave can be expressed as:
\begin{equation}
\Psi =\left( \sqrt{\left( 1-\frac{2\pi ~{Im}A_{0}}{k~d_{y}}\right)
^{2}+\left( \frac{2\pi ~{Re}A_{0}}{k~d_{y}}\right) ^{2}}\right)^m
e^{ikz} e^{i\varphi m},
\label{m_planes}
\end{equation}
where $\varphi =arctg\left( \frac{\frac{2\pi ~Re A_{0}}{k~d_{y}}}{1-\frac{%
2\pi ~Im A_{0}}{k~d_{y}}}\right)$, $m=\frac{z}{d_z}$ inside the structure formed by threads. This expression can be easily converted to the form
$\Psi =e^{iknz}$, where $n$ is the refraction index defined as
\begin{eqnarray}
n& = & n^{\prime }+in^{\prime \prime }=\\
& = & \left( 1+\frac{\lambda }{2\pi d_{z}}%
Arctg\left( \frac{\frac{\lambda }{d_{y}}{Re}A_{0}}{1-\frac{\lambda }{%
d_{y}}{Im}A_{0}}\right) \right) -i\frac{\lambda }{2\pi d_{z}}\ln \left(
\sqrt{\left( \frac{\lambda }{d_{y}}{Re}A_{0}\right) ^{2}+\left( 1-\frac{%
\lambda }{d_{y}}{Im}A_{0}\right) ^{2}}\right). \nonumber
\label{n_exact}
\end{eqnarray}
here $\lambda=\frac{2 \pi}{k}$ is used.

If $Re A_{0}, ~Im A_{0} \ll 1$ then (\ref{n_exact}) can be
expressed as:
\begin{equation}
n=1+\frac{2 \pi}{d_y d_z k^2}A_0.
\label{n_Bar}
\end{equation}

Radiation frequencies of our interest is  $\nu \ge 10$ GHz. In this frequency range
skin depth $\delta$ is about 1 micron for the most of metals (for example, $\delta_{Cu}=0.66$ $\mu$m,
$\delta_{Al}=0.8$ $\mu$m, $\delta_{W}=1.16$ $\mu$m and so on). Thus, in this frequency range the metallic threads can be
considered as perfect conducting.

From the analysis \cite{Nikolsky} follows that the amplitude $A_0$ for the
perfect conducting cylinder for polarization of
the electromagnetic wave
parallel to the cylinder
axis can be expressed as:
\begin{equation}
A_{0\left( \parallel \right) }=\frac{1}{\pi }\frac{J_{0}\left( kR\right)
N_{0}\left( kR\right) }{J_{0}^{2}\left( kR\right) +N_{0}^{2}\left( kR\right)
}+i\frac{1}{\pi }\frac{J_{0}^{2}\left( kR\right) }{J_{0}^{2}\left( kR\right)
+N_{0}^{2}\left( kR\right) }
\label{A0_par}
\end{equation}
Amplitude $A_0$ for the perfect conducting cylinder for polarization of
the electromagnetic wave  perpendicular
to the cylinder axis is as follows \cite{Nikolsky}:
\begin{equation}
A_{0\left( \perp \right) }=\frac{1}{\pi }\frac{J_{0}^{\prime }\left(
kR\right) N_{0}^{\prime }\left( kR\right) }{J_{0}^{\prime 2}\left( kR\right)
+N_{0}^{\prime 2}\left( kR\right) }+i\frac{1}{\pi }\frac{J_{0}^{\prime
2}\left( kR\right) }{J_{0}^{\prime 2}\left( kR\right) +N_{0}^{\prime
2}\left( kR\right) },
\label{A0_perp}
\end{equation}
where $R$ is the cylinder (thread) radius, $J_{0}, N_{0}, J_{0}^{\prime}$ and $N_{0}^{\prime}$
are the Bessel and Neumann functions and their derivatives, respectively. Using the
asymptotic values for these functions for $kR \ll 1$ one can obtain:
\begin{eqnarray}
\begin{array}{l}
J_{0}\left( x\rightarrow 0\right) \approx 1,~
N_{0}\left( x\rightarrow 0\right) \approx -\frac{2}{\pi }\ln \frac{2}{%
1.781\cdot x},\\
J_{0}^{\prime }\left( x\rightarrow 0\right) =-J_{1}\approx
-\frac{x}{2},~
N_{0}^{\prime }\left( x\rightarrow 0\right) =-N_{1}\approx -\frac{2}{\pi }%
\frac{1}{x}.
\end{array}
\end{eqnarray}

Let us consider a particular example. Suppose  radiation frequency $\nu=10$ GHz and the thread radius
$R=0.1$ mm, then
\begin{eqnarray}
& &ReA_{0\left( \parallel \right)} \approx -0.1087,~ImA_{0\left( \parallel \right)} \approx 0.0429, \\
\label{A0reimpar}
& &ReA_{0\left( \perp \right)} \approx -0.00011,~ImA_{0\left( \perp \right)} \approx 3.78 \cdot 10^{-8}
\label{A0reimperp}
\end{eqnarray}
and
\begin{eqnarray}
& &n_{\parallel}=0.8984+i \cdot 0.043 , \\
\label{n_par} & &n_{\perp}=0.9998+i \cdot 3.37 \cdot 10^{-9}
\label{n_perp}
\end{eqnarray}
Such values for $n$ provides to conclude that, in contrast to a solid metal,
an electromagnetic wave falling on the described "grid" volume structure
is not absorbed on the skin depth, but passes through the "grid" damping
in accordance its polarization.

The electromagnetic wave with polarization parallel to the thread
axis is strongly absorbed while passing through the structure.
Absorption for the wave with polarization perpendicular to the
thread axis is weak.

Difference in $n_{\parallel}$ and $n_{\perp}$ indicates that the
system own optical anysotropy (i.e. possesses birefringence and
dichroism). To escape this anysotropy we can use alternating the
treads position in grid: threads in each layer are orthogonal to
the threads in previous and following layer.

And one more important note: When an electron beam passes through
volume grid The pulse of transition radiation appears due to large
value of the refraction index

The values
$ReA_{0\left( \parallel \right)}$ and $ImA_{0\left( \parallel \right)}$ are quite large and
for polarization parallel to the thread axis the exact expression (\ref{m_planes},\ref{n_exact}) should be
used. Moreover, in all calculations we should carefully check whether the condition $\left\vert
n-1\right\vert \ll 1$ is fullfilled. If no, then we should use more strict description of volume
structure and consider rescattering of the wave by different threads.

In this case in contrast to (\ref{sum_cylinder}) the
electromagnetic wave is described by (the wave with polarization
along the thread axes is considered):
\begin{equation}
\Psi(\rho)=e^{ikz}+\Sigma_{m} F_m H_{0}^{(1)} (k \left\vert
\overrightarrow{\rho }-\overrightarrow{\rho }_{m}\right\vert),
\label{sum_ncylinder}
\end{equation}
where $F_m$ is the effective scattering amplitude defined by
\begin{equation}
F_m=a_0 e^{ikz_m}+a_0\Sigma_{n \ne m} F_n H_{0}^{(1)} (k
\left\vert \overrightarrow{\rho }-\overrightarrow{\rho
}_{n}\right\vert), \label{F_m}
\end{equation}
%
%
%
%
Let us consider unregulated set of threads.
According to the system of equations
(\ref{sum_ncylinder},\ref{F_m}) to find $F_m$ it is necessary to
solve the system of algebraic equations.

 Let us consider first the long-wave case ($kd \ll 1$) to obtain
 the approximate solution; the sum in (\ref{F_m}) could be
 replaced by the integral:
\begin{equation}
F_m(\rho)=a_0 \left\{ e^{ikz}+
 \frac{1}{\Omega_2} \int_V F_m(\rho^{\prime}) H_{0}^{(1)}
(k \left\vert \overrightarrow{\rho}-\overrightarrow{\rho
}^{\prime}\right\vert) d^2 \rho^{\prime} \right\}, \label{n0}
\end{equation}
where the area $V$ includes all scatterers with the exception of
that located in the point $\rho$ (this area $V$ being supplied
with the small area $\Delta V$, surrounding the point $\rho$,
gives us the whole area)

 If note here that
\begin{equation}
 e^{ikz}+
 \frac{1}{\Omega_2} \int F_m(\rho^{\prime}) H_{0}^{(1)}
(k \left\vert \overrightarrow{\rho}-\overrightarrow{\rho
}^{\prime}\right\vert) d^2 \rho^{\prime}=\Psi(\rho), \label{n1}
\end{equation}
and present the integral over the area $V$ as the difference of
integral over the whole area and the integral over the area
$\Delta V$ we can get
\begin{equation}
F_m(\rho)=a_0 \left\{ \Psi(\rho) -
 \frac{1}{\Omega_2} \int_{\Delta V} F_m(\rho^{\prime}) H_{0}^{(1)}
(k \left\vert \overrightarrow{\rho}-\overrightarrow{\rho
}^{\prime}\right\vert) d^2 \rho^{\prime} \right\}, \label{n2}
\end{equation}
as the area $\Delta V$ is small surrounding of the point $\rho$
and $F_m(\rho^{\prime})$ does not change significantly within this
area, the values $F_m(\rho^{\prime})$ in the integral over $\Delta
V$ could be replaced by $F_m(\rho)$. This provides to write
\begin{equation}
F_m(\rho)=a_0 \Psi(\rho) -
 F_m(\rho)\frac{a_0}{\Omega_2} \int_{\Delta V} H_{0}^{(1)}
(k \left\vert \overrightarrow{\rho}-\overrightarrow{\rho
}^{\prime}\right\vert) d^2 \rho^{\prime} , \label{n3}
\end{equation}
Therefore
\begin{equation}
 F_m(\rho)=\frac{a_0}{1+\frac{a_0}{\Omega_2} \int_{\Delta
V} H_{0}^{(1)} (k \left\vert
\overrightarrow{\rho}-\overrightarrow{\rho }^{\prime}\right\vert)
d^2 \rho^{\prime}} \Psi(\rho) , \label{n4}
\end{equation}
and the equation for $\Psi$ looks as follows
\begin{equation}
\Psi(\rho)= e^{ikz}+
 \frac{B}{\Omega_2} \int_V  H_{0}^{(1)}
(k \left\vert \overrightarrow{\rho}-\overrightarrow{\rho
}^{\prime}\right\vert) \Psi(\rho^{\prime})
d^2 \rho^{\prime} ,
\label{n5}
\end{equation}
or
\begin{equation}
(\Delta+k^2)\Psi(\rho)=
 \frac{4 i B}{\Omega_2} \Psi(\rho)
\label{n6}
\end{equation}
When searching $\Psi(\rho) \sim e^{iq\rho}$ we get
\begin{equation}
-q^2+k^2=
 \frac{4 i B}{\Omega_2},
\label{n6}
\end{equation}
after introducing parameter $b_0=B/a_0$ and remembering $a_0=i \pi
A_0$ (\ref{n6}) appears
\begin{equation}
q^2=k^2
 +\frac{4 \pi A_0 b_0}{\Omega_2},
\label{n6}
\end{equation}
where $b_0=\frac{1} {1+\frac{a_0}{\Omega_2}\int_{\Delta V}
H_{0}^{(1)} (k \left\vert
\overrightarrow{\rho}-\overrightarrow{\rho }^{\prime}\right\vert)
d^2 \rho^{\prime}}$.

To find $b_0$ the integral should be calculated:
\begin{equation}
\frac{1}{\Omega_2}\int_{\Delta V} H_{0}^{(1)} (k \left\vert
\overrightarrow{\rho}-\overrightarrow{\rho }^{\prime}\right\vert)
d^2
\rho^{\prime}=(1+i\frac{2}{\pi}C)+i\frac{2}{\pi}\ln{\frac{kR}{2}},
\label{n7}
\end{equation}

Thus, the refraction index is as follows:
\begin{equation}
n^2=1+ \frac{4 \pi A_0}{\Omega_2 k^2} \frac{1}{1+a_0
((1+i\frac{2}{\pi}C)+i\frac{2}{\pi}\ln{\frac{kR}{2}})}, \label{n8}
\end{equation}
or rewriting with $a_0=i \pi A_0$
\begin{equation}
n^2=1+ \frac{4 \pi A_0}{\Omega_2 k^2}~ \frac{1}{1+i \pi A_0 -2C
A_0-2 A_0 \ln{\frac{kR}{2}}}, \label{n8}
\end{equation}
here $C=0.5772$ is the Eiler constant.

Let us consider the regular structure (the volume "grid") built
from treads. Scattering by a thread is described by:
\begin{equation}
\Psi(\rho)=e^{ikz}+a_0 H_0^{(1)} (k \left\vert
\overrightarrow{\rho}-\overrightarrow{\rho
}^{\prime}\right\vert)e^{ikz_n} \label{n9}
\end{equation}
When $z \rho \rightarrow z_n \rho_n$  the wavefunction $\Psi$ can
be expressed:
\begin{equation}
\Psi(\rho)=e^{ikz_n}+a_0
 \{1+i \frac{2}{\pi} C +i \frac{2}{\pi} \ln \frac{k |\vec{\rho}-\vec{\rho}_n|}{2} \}
e^{ikz_n} \label{n9a}
\end{equation}
Introducing $\varphi=1+ a_0(1+i \frac{2}{\pi} C)$ that gathers
non-divergent terms, we can rewrite this expression as follows
\begin{equation}
\Psi(\rho)=e^{ikz_n} \varphi
 \{1+ \frac{a_0}
 {1+ a_0(1+i \frac{2}{\pi} C)} i \frac{2}{\pi} \ln \frac{k |\vec{\rho}-\vec{\rho}_n|}{2} \}
e^{ikz_n} \label{n9a}
\end{equation}

Using the similar reasoning for many scatterers (considering $z
\rightarrow z_n$) we can obtain for the wavefunction:
\begin{eqnarray}
\begin{array}{l}
\Psi(\rho)=e^{ikz_n} + (...+ ..._{non-divergent~terms}... + ...) +
F_n
  i \frac{2}{\pi} \ln \frac{k |\vec{\rho}-\vec{\rho}_n|}{2}
e^{ikz_n} = \\
=e^{ikz_n} \varphi_n \{ 1 + \frac{F_n}{\varphi_n} i \frac{2}{\pi}
\ln \frac{k |\vec{\rho}-\vec{\rho}_n|}{2}\},
\end{array}
 \label{n9a}
\end{eqnarray}
where $\frac{F_n}{\varphi_n}=\frac{a_0}{1+ a_0(1+i \frac{2}{\pi}
C)}$.

The solution in a volume "grid" (an artificial crystal) could be
presented in the form:
\begin{equation}
\Psi(\vec{\rho})=\chi(\vec{\rho})
e^{i\vec{k}^{\prime}\vec{\rho}},~ \vec{\rho}=(y,z) \label{n10}
\end{equation}
The equation for the wavefunction
\begin{equation}
(\Delta+k^2)\Psi(\vec{\rho})= 4 i F \sum_m
e^{i\vec{k}^{\prime}\vec{\rho}}
\delta(\overrightarrow{\rho}-\overrightarrow{\rho}_m), \label{n11}
\end{equation}
where $F$ is an amplitude, provides to get the equation for
$\chi(\vec{\rho})$:
\begin{equation}
\Delta \chi(\vec{\rho})+2i
\vec{k}^{\prime}\vec{\nabla}\chi(\vec{\rho})-({k^{\prime}}^2-k^2)\chi(\vec{\rho})=4
i F \sum_m e^{i\vec{k}^{\prime}\vec{\rho}}
\delta(\overrightarrow{\rho}-\overrightarrow{\rho}_m), \label{n12}
\end{equation}
where $\chi(\vec{\rho})$ can be presented as a sum over the
reciprocal lattice vectors $\vec{\tau}$
\begin{equation}
\chi(\vec{\rho})=\sum_{\tau} c_{\tau} e^{i \vec{\tau}\vec{\rho}},
\label{n13}
\end{equation}
Therefore, the wavefunction can be expressed
\begin{equation}
\Psi(\vec{\rho})=-\frac{4 i F}{\Omega_2}\sum_{\tau} \frac{e^{i
(\vec{k}^{\prime}+\vec{\tau})\vec{\rho}}}{(\vec{k}^{\prime}+\vec{\tau})^2-k^2}.
\label{n14}
\end{equation}
At the limit $\rho \rightarrow 0$
\begin{equation}
\Psi - F i \frac{2}{\pi} \ln \frac{k \rho}{2}=\varphi
\end{equation}
Substituting the expression (\ref{n14})
\begin{equation}
-\frac{4 i F}{\Omega_2}\sum_{\tau} \frac{e^{i
(\vec{k}^{\prime}+\vec{\tau})\vec{\rho}}}{(\vec{k}^{\prime}+\vec{\tau})^2-k^2}-F
i \frac{2}{\pi} \ln \frac{k \rho}{2}=\varphi
\label{n14a}
\end{equation}
i.e.
\begin{equation}
\frac{4 i }{\Omega_2}\sum_{\tau} \frac{e^{i
(\vec{k}^{\prime}+\vec{\tau})\vec{\rho}}}{(\vec{k}^{\prime}+\vec{\tau})^2-k^2}+
i \frac{2}{\pi} \ln \frac{k
\rho}{2}=-\frac{\varphi}{F}=\frac{F_n}{\varphi_n}=\frac{a_0}{1+
a_0(1+i \frac{2}{\pi} C)}
 \label{n14a}
\end{equation}
\begin{equation}
{{k}^{\prime}}^2=k^2+\frac{4
\pi}{\Omega_2}\frac{A_0}{1+a_0(1+i\frac{2}{\pi}C)}=k^2+\frac{4
\pi}{\Omega_2}\frac{A_0}{1+i \pi A_0-2CA_0}, \label{n15}
\end{equation}

Therefore, the index of refraction is
\begin{equation}
{n}^2=1+\frac{4 \pi A_0}{\Omega_2 k^2}~\frac{1}{1+i \pi
A_0-2CA_0}, \label{n15}
\end{equation}


Then for the same example
($\nu=10$ GHz, $R=0.1$ mm) we obtain:
\begin{eqnarray}
& &n_{\parallel}=0.779+i \cdot 1.217 \cdot 10^{-19} , \\
\label{n_par_new} & &n_{\perp}=0.99987 - i \cdot 1.3464 \cdot
10^{-23} \label{n_perp_new}
\end{eqnarray}
(compare this with (12,13)).

Rescattering effects significantly change the index of refraction and its imaginary part
appears noticeably reduced.

\section{Sketch theory of VFEL lasing using electron beam radiation in a volume "grid" resonator}
Maxwell equations and motion equations in this case
\begin{eqnarray}
\begin{array}{l}
rot \vec{H}=\frac{1}{c}\frac{\partial \vec{D}}{\partial t}+\frac{4
\pi}{c} \vec{j},~\vec{D}(\vec{r,t})=\int_{-\infty}^{\infty}
\varepsilon(\vec{r},t-t^{\prime})\vec{E}(\vec{r},t^{\prime})
dt^{\prime},\\
rot \vec{E}=-\frac{1}{c}\frac{\partial \vec{H}}{\partial t},~ div
\vec{D}=4 \pi \rho,~\frac{\partial \rho}{\partial t}+div
\vec{j}=0, \\
\end{array}
\end{eqnarray}
here $\varepsilon(\vec{r},t-t^{\prime})<0$ at $t<t^{\prime}$,
$D_i(\vec{r},t^{\prime})=\int \varepsilon
(\vec{r},t-t^{\prime})E_l(\vec{r},t^{\prime})dt^{\prime}$ and,
therefore, $D_i(\vec{r},\omega)=\varepsilon_{il}
(\vec{r},\omega)E_l(\vec{r},\omega)$. Combining the above equation
we obtain:
\begin{eqnarray}
-\Delta
\vec{E}+\vec{\nabla}(\vec{\nabla}\vec{E})+\frac{1}{c^2}\frac{\partial^2
\vec{D}}{\partial t^2}=-\frac{4 \pi}{c^2} \frac{\partial
\vec{j}}{\partial t}
\end{eqnarray}
after making the Fourier transformation:
\begin{eqnarray}
\begin{array}{l}
rot rot \vec{E}(\vec{r},\omega)-\frac{\omega^2}{c^2}\varepsilon
(\vec{r},\omega) \vec{E}(\vec{r},\omega=\frac{4 \pi i
\omega}{c^2} \vec{j}(\vec{r},\omega) \\
div~ \varepsilon(\vec{r},\omega)\vec{E}(\vec{r},\omega)=4 \pi
\rho(\vec{r},\omega),\\
-i\omega \rho(\vec{r},\omega)+ div \vec{j}(\vec{r},\omega)=0,\\
\vec{j}(\vec{r},t)=e
\sum_{\alpha}\vec{v}_{\alpha}(t)\delta(\vec{r}-\vec{r}_{\alpha}(t)),~
\rho(\vec{r},t)=e
\sum_{\alpha}\delta(\vec{r}-\vec{r}_{\alpha}(t)), \\
\frac{d \vec{v}_{\alpha}}{dt}=\frac{e}{m \gamma}\left\{
 \vec{E}(\vec{r}_{\alpha}(t),t)+\frac{1}{c} [
\vec{v}_{\alpha}(t) \times \vec{H}(\vec{r}_{\alpha}(t),t)
 ]-
 \frac{\vec{v}_{\alpha}}{c^2}(\vec{v}_{\alpha}(t)\vec{E}(\vec{r}_{\alpha}(t),t))
\right\}
\end{array}
\label{sys1}
\end{eqnarray}
where $\gamma=(1-\frac{v_{\alpha}^2}{c^2})^{-\frac{1}{2}}$.

Applying the method of slow-varying amplitudes the solution for
this system can be expressed as
\begin{equation}
\vec{E}(\vec{r},t)= \vec{e}_1 A_1(\vec{r},t)e^{i(\vec{k}_1
\vec{r}-\omega t)}+\vec{e}_2 A_2(\vec{r},t)e^{i(\vec{k}_2
\vec{r}-\omega t)}, \label{E}
\end{equation}
$\vec{k}_2=\vec{k}_1+\vec{\tau}$.
Substituting (\ref{E}) to the exact system of equations and
collecting the quick-oscillating terms we obtain the system:
\begin{eqnarray}
\begin{array}{l}
2 i \vec{k}_1 \vec{\nabla}A_1 (\vec{r},t) - k_1^2 A_1(\vec{r},t)
+\frac{\omega^2}{c^2} \varepsilon^0(\omega) A_1(\vec{r},t)+i
\frac{1}{c^2}\frac{\partial \omega^2
\varepsilon^0(\omega)}{\partial \omega} \frac{\partial
A_1(\vec{r},t) }{\partial t}+ \\
+\frac{\omega^2}{c^2}\varepsilon^{-\tau}(\omega) A_2 (\vec{r},t)+
i \frac{1}{c^2} \frac{\partial \omega^2
\varepsilon^{-\tau}(\omega)}{\partial \omega}\frac{\partial
A_2(\vec{r},t) }{\partial t}=J_1, \\
2 i \vec{k}_2 \vec{\nabla}A_2 (\vec{r},t) - k_2^2 A_2(\vec{r},t)
+\frac{\omega^2}{c^2} \varepsilon^0(\omega) A_2(\vec{r},t)+
 i \frac{1}{c^2}\frac{\partial \omega^2
\varepsilon^0(\omega)}{\partial \omega} \frac{\partial
A_2(\vec{r},t) }{\partial t}+ \\
+\frac{\omega^2}{c^2}\varepsilon^{\tau}(\omega) A_1 (\vec{r},t)+ i
\frac{1}{c^2} \frac{\partial \omega^2
\varepsilon^{\tau}(\omega)}{\partial \omega}\frac{\partial
A_1(\vec{r},t) }{\partial t}=J_2,
\end{array}
\label{sys2}
\end{eqnarray}
$J_1$, $J_2$ are the currents, their explicit expressions can be
found in \cite{Batrakov+Sytova}.

The system (\ref{sys2}) includes terms describing wave dispersion,
if omit these terms we get the system analyzed in the foregoing
paper \cite{Batrakov+Sytova}.

Consideration of scattering by a diffraction grating in a
waveguide requires $\vec{E}(\vec{r},t)$ expansion over the
waveguide eigenfunctions $\vec{Y}_{\lambda mn}(\vec{r}_{\perp})$
($\vec{r}_{\perp}=(x,y)$, the waveguide axes is parallel to the
axes $z$), which meet the equation:
\begin{equation}
\Delta_{\perp} \vec{Y}_{\lambda mn}+
({\varkappa_{mn}^{~\lambda}})^2  \vec{Y}_{\lambda mn}=0.
\end{equation}
Expanding $\vec{E}(\vec{r},t)=\vec{E}(\vec{r}_{\perp},z,t)$ over
$\vec{Y}_{\lambda mn}(\vec{r}_{\perp})$
\begin{equation}
\vec{E}(\vec{r}_{\perp},z,t)=\sum_{\lambda mn} C_{\lambda mn}(z,t)
\vec{Y}_{\lambda mn}(\vec{r}_{\perp})
\end{equation}
and considering the waveguide with a diffraction grating in vacuum
\begin{equation}
D_i (\vec{r}, \omega)= (\delta_{il}+\chi_{il}(\vec{r},
\omega))E_l(\vec{r}, \omega)= E_i(\vec{r},
\omega)+\chi_{il}(\vec{r}, \omega) E_l(\vec{r}, \omega)
\end{equation}
we can write
\begin{eqnarray}
\begin{array}{l}
\Delta \vec{E} - \vec{\nabla}(div \vec{E}) -
\frac{1}{c^2}\frac{\partial^2}{\partial t^2} \int
\hat{\varepsilon}(t-t^{\prime})
\vec{E}(t^{\prime})dt^{\prime}=\frac{4 \pi}{c^2} \frac{\partial
\vec{j}}{\partial t},
\end{array}
\end{eqnarray}
or
\begin{eqnarray}
\begin{array}{l}
\Delta \vec{E} - \frac{1}{c^2}\frac{\partial^2}{\partial t^2} \int
\hat{\varepsilon}(t-t^{\prime}) (\vec{E}(t^{\prime}) -
\vec{\nabla}(div \hat{\chi}\vec{E}))dt^{\prime}=\frac{4 \pi}{c^2}
\frac{\partial \vec{j}}{\partial t}+ 4 \pi \vec{\nabla} \rho,
\end{array}
\end{eqnarray}
or
\begin{eqnarray}
\begin{array}{l}
\Delta \vec{E} - \frac{1}{c^2}\frac{\partial^2 \vec{E}}{\partial
t^2}- \frac{1}{c^2}\frac{\partial^2}{\partial t^2} \int
\hat{\chi}(t-t^{\prime})\vec{E}(t^{\prime}) - \vec{\nabla}(div
\hat{\chi}\vec{E}))dt^{\prime}=\frac{4 \pi}{c^2} \frac{\partial
\vec{j}}{\partial t}+ 4 \pi \vec{\nabla} \rho.
\end{array}
\end{eqnarray}
After expansion
\begin{eqnarray}
\begin{array}{l}
\frac{\partial^2 C_{\lambda mn}(z,t)}{\partial z^2}-\frac{1}{c^2}
\frac{\partial^2 C_{\lambda mn}(z,t)}{\partial
t^2}-({\varkappa_{mn}^{~\lambda}})^2 C_{\lambda mn}(z,t)-\\
-\frac{1}{c^2}\frac{\partial^2}{\partial t^2}
 {\int \vec{Y}_{\lambda mn}^{*}(\vec{r}_{\perp})} \hat{\chi} (t-t^{\prime})
 \sum_{{\lambda}^{\prime}m^{\prime}n^{\prime}} C_{{\lambda}^{\prime}m^{\prime}n^{\prime}}(z,t^{\prime})
  \vec{Y}_{{\lambda}^{\prime}m^{\prime}n^{\prime}}(\vec{r}_{\perp})
  d^2 {r}_{\perp}+\\
 +\vec{Y}_{\lambda mn}^{*}(\vec{r}_{\perp})vec{\nabla}(div \vec{\chi}
  \sum_{{\lambda}^{\prime}m^{\prime}n^{\prime}}
  C_{{\lambda}^{\prime}m^{\prime}n^{\prime}}(z,t^{\prime}))
  \vec{Y}_{{\lambda}^{\prime}m^{\prime}n^{\prime}}(\vec{r}_{\perp})  d^2 {r}_{\perp}=\\
=\frac{4 \pi}{c^2}\frac{\partial}{\partial t} {\int
\vec{Y}_{\lambda mn}^{*}(\vec{r}_{\perp}) \vec{j}_{\lambda
mn}}d^2({r}_{\perp})+4 \pi {\int \vec{Y}_{\lambda
mn}^{*}(\vec{r}_{\perp}) \vec{\nabla} \rho d^2({r}_{\perp})}
\end{array}
\end{eqnarray}
Applying the method of slow variable amplitudes we can obtain the
system of equations describing the waves excited in the system:
$C_{\lambda mn}(z,t)=A_{\lambda mn}(z,t) e^{i(\varkappa_{\lambda
mn} z- \omega t)}$ (here $\varkappa_{\lambda mn}$ and $\omega$
corresponds to the waveguide without a diffraction grating).

In general case different modes are separated, but grating
rotation could mix different modes \cite{FEL2002} (similar waves
mixing in the vicinity of Bragg condition). To describe this
process the equations for the mixing modes should be solved
conjointly.

\section{Conclusion}

In the present paper the electrodynamical properties of a volume
resonator that is formed by a periodic structure built from the
metallic threads inside a rectangular waveguide is considered.
Peculiarities of passing of electromagnetic waves with different
polarizations through such volume resonator are discussed. If in
the periodic structure built from the metallic threads diffraction
conditions are available, then analysis shows that in this system
the effect of anomalous transmission for electromagnetic waves
could appear similarly to the Bormann effect well-known in the
dynamical diffraction theory of X-rays.

\end{document}